\documentclass[aps,prl,showpacs,twocolumn,groupedaddress,superscriptaddress]
{revtex4}
\usepackage{graphicx} 
\usepackage{dcolumn} 
\usepackage{longtable} 
\usepackage{amssymb} 
\usepackage{amsmath} 
\usepackage{amsfonts} 
\usepackage{slashed} 
\usepackage[dvipsnames]{xcolor} 

\newcommand{\bs}[1]{\boldsymbol{#1}} 
\newcommand{\nopieft}{\mbox{$\slashed{\pi}$EFT}} 
 
\newcommand{\Lag}{{\cal L}} 
\newcommand{\be}{\begin{equation}} 
\newcommand{\ee}{\end{equation}} 
\newcommand{\rvec}{{\bs{r}}} 
\newcommand{\xvec}{{\bs{x}}}

\newcommand{\half}{\frac{1}{2}} 

\begin{document}

\title{Resolving the $\Lambda$ Hypernuclear Overbinding Problem in Pionless 
Effective Field Theory}

\author{L.~Contessi} 
\affiliation{Racah Institute of Physics, The Hebrew University, 
Jerusalem 91904, Israel}
\author{N.~Barnea} 
\affiliation{Racah Institute of Physics, The Hebrew University, 
Jerusalem 91904, Israel}
\author{A.~Gal}\thanks{avragal@savion.huji.ac.il}\affiliation{Racah Institute 
of Physics, The Hebrew University, Jerusalem 91904, Israel} 

\date{\today}

\begin{abstract} 

We address the $\Lambda$-hypernuclear `overbinding problem' in light 
hypernuclei which stands for a 1--3~MeV excessive $\Lambda$ separation 
energy calculated in $_{\Lambda}^5$He. This problem arises in most few-body 
calculations that reproduce ground-state $\Lambda$ separation energies in 
the lighter $\Lambda$ hypernuclei within various hyperon-nucleon interaction 
models. Recent pionless effective field theory (\nopieft) nuclear few-body 
calculations are extended in this work to $\Lambda$ hypernuclei. At leading 
order, the $\Lambda N$ low-energy constants are associated with $\Lambda N$ 
scattering lengths and the $\Lambda NN$ low-energy constants are fitted to 
$\Lambda$ separation energies ($B_{\Lambda}^{\rm exp}$) for $A\leq 4$. The 
resulting \nopieft~interaction reproduces in few-body stochastic variational 
method calculations the reported value $B_{\Lambda}^{\rm exp}(_{\Lambda}^5
$He)=3.12$\pm$0.02~MeV within a fraction of MeV over a broad range of 
\nopieft~cut-off parameters. Possible consequences and extensions to 
heavier hypernuclei and to neutron-star matter are discussed. 

\end{abstract}

\maketitle

\noindent 
{\bf Introduction.}~~$\Lambda$ hypernuclei provide extension of atomic nuclei 
into the strangeness sector of hadronic matter~\cite{GHM16}. Experimental 
data on $\Lambda$ hypernuclei are poorer, unfortunately, in both quantity and 
quality than the data available on normal nuclei. Nevertheless, the few dozen 
of $\Lambda$ separation energies $B_{\Lambda}^{\rm exp}$ determined across 
the periodic table, mostly for hypernuclear ground states, provide a useful 
test ground for the role of strangeness in dense hadronic matter, for example 
in neutron stars~\cite{LLGP15}. Particularly meaningful tests of 
hyperon-nucleon ($YN$) strong-interaction models are possible in light 
$\Lambda$ hypernuclei, $A\leq 5$, where precise few-body {\it ab initio} 
calculations are feasible~\cite{Nogga13}. 

\begin{table}[ht!] 
\caption{Ground-state $\Lambda$ separation energies $B_{\Lambda}$ and 
excitation energies $E_x$ (in MeV) from several few-body calculations 
of $s$-shell $\Lambda$ hypernuclei; see the text. Charge symmetry breaking 
(CSB) is included in the $_\Lambda^4$H results from Ref.~\cite{GG16}.}  
\begin{ruledtabular} 
\begin{tabular}{ccccc} 
  & $B_{\Lambda}(_\Lambda^3$H) & $B_{\Lambda}(_\Lambda^4$H$_{\rm gs}$) 
& $E_x(_\Lambda^4$H$_{\rm exc}$) & $B_{\Lambda}(_\Lambda^5$He)  \\
Exp. & 0.13(5)~\cite{Davis05} & 2.16(8)~\cite{MAMI16} & 1.09(2)~\cite{E13} & 
3.12(2)~\cite{Davis05}  \\
\hline 
DHT~\cite{DHT72} & 0.10 & 2.24 & 0.36 & $\geq$5.16  \\ 
AFDMCa & -- & 1.97(11)~\cite{LPG14} & -- & 5.1(1)~\cite{LGP13}  \\ 
AFDMCb & $-$1.2(2)~\cite{LPG14} & 1.07(8)~\cite{LPG14} & -- & 
3.22(14)~\cite{LPG14}  \\ 
$\chi$EFTa & 0.11(1)~\cite{Wirth14} & 2.31(3)~\cite{GG16} & 
0.95(15)~\cite{GG16} & 5.82(2)~\cite{WR18}  \\  
$\chi$EFTb & -- & 2.13(3)~\cite{GG16} & 1.39(15)~\cite{GG16} & 
4.43(2)~\cite{WR18}  \\ 
\end{tabular} 
\end{ruledtabular} 
\label{tab:over} 
\end{table} 

The $\Lambda N$ interaction is not sufficiently strong to bind 
two-body systems. Hypernuclear binding starts with the weakly bound 
$_{\Lambda}^3$H($I$=0,$J^P$=${\frac{1}{2}}^+$) hypernucleus. No other $A$=3 
hypernuclear level has ever been firmly established. The $A$=4 isodoublet 
hypernuclei ($_{\Lambda}^4$H, $_{\Lambda}^4$He) each have two bound states, 
$0^+_{\rm gs}$ and $1^+_{\rm exc}$. The hypernuclear $s$ shell ends with 
a single $_{\Lambda}^5$He($I$=0,$J^P$=${\frac{1}{2}}^+$) level. 
Table~\ref{tab:over} demonstrates in chronological order the extent to which 
several representative few-body calculations overbind $_{\Lambda}^5$He while 
reproducing the $B_{\Lambda}$ values of all other $s$-shell hypernuclear 
levels. This is known as the `overbinding problem' in light $\Lambda$ 
hypernuclei since the 1972 work by Dalitz, Herndon and Tang (DHT)~\cite{DHT72} 
who used a phenomenological $\Lambda N$+$\Lambda NN$ interaction model. 
The other, recent calculations listed in the table use the following 
methodologies: 
\newline 
(i) Auxiliary-field diffusion Monte Carlo (AFDMC) techniques within a $\Lambda 
N$+$\Lambda NN$ Urbana-type interaction model dating back to Bodmer, Usmani 
and Carlson~\cite{BUC84}. Note that, while version AFDMCb~\cite{LPG14} 
reproduces $B_{\Lambda}^{\rm exp}(_{\Lambda}^5$He) as a prerequisite to 
resolving the `hyperon puzzle' in neutron-star matter~\cite{LLGP15}, it 
underbinds the lighter $s$-shell hypernuclei by about 1~MeV each and, thus, 
does not resolve the overbinding problem as defined here. A revision of this 
work~\cite{LP17} suggests that by modifying some of the $\Lambda NN$ strength 
parameters it is possible to avoid the underbinding. 
\newline 
(ii) No-core shell-model techniques within a leading-order (LO) chiral 
effective field theory ($\chi$EFT) $YN$ interaction model, with momentum 
cut-off values of 600 (a) and 700 (b)~MeV/c, in which three-body $\Lambda NN$ 
terms are induced through $\Lambda N \leftrightarrow \Sigma N$ coupling. The 
$_{\Lambda}^5$He $\chi$EFT results listed here were obtained by employing 
a similarity renormalization group transformation~\cite{WR18}, reducing 
the model-space dimension in order to enhance the poor convergence met in 
using bare $YN$ interactions~\cite{WGNR18}. No $\chi$EFT calculations have 
been reported yet for $_{\Lambda}^5$He at next-to-leading order (NLO). 

Excluding calculations using an uncontrolled number of interaction terms, the 
only published few-body calculations claiming to have solved the overbinding 
problem are those by Nemura et al.~\cite{NAS02}. However, it was realized by 
Nogga, Kamada and Gl\"{o}ckle~\cite{NKG02} that a more faithful reproduction 
of the Nijmegen soft-core (NSC) meson-exchange potentials used in these 
calculations in fact {\it underbinds} appreciably the $A$=4 hypernuclei. Thus, 
the overbinding problem is still alive and kicking, with $_{\Lambda}^5$He 
overbound by 1--3~MeV in the recent few-body calculations listed in 
Table~\ref{tab:over}. 

The present work reports on few-body stochastic variational method (SVM) 
precise calculations of $s$-shell hypernuclei, using Hamiltonians constructed 
at LO in a pionless effective field theory (\nopieft) approach. This is 
accomplished by extending a purely nuclear \nopieft~Hamiltonian used in 
few-nucleon calculations, first reported in Refs.~\cite{Kol99,BHK00} and more 
recently also in lattice-nuclei calculations~\cite{BCG15,KBG15,CLP17,KPDB17}, 
to include $\Lambda$ hyperons. With $\Lambda N$ one-pion exchange (OPE) 
forbidden by isospin invariance, the \nopieft~breakup scale is 2$m_{\pi}$, 
remarkably close to the threshold value $p_{\Lambda N}^{\rm th}\approx 283
$~MeV/c for exciting $\Sigma N$ pairs in $\pi$EFT approaches~\cite{Haiden07}. 
A typical momentum scale $Q$ in $_{\Lambda}^5$He is $p_{\Lambda}\approx\sqrt{
2M_{\Lambda}B_{\Lambda}}=83$~MeV/c, suggesting a \nopieft~expansion parameter 
$(Q/2m_{\pi})\approx 0.3$ for $s$-shell hypernuclei. This implies a $\nopieft$ 
LO accuracy of the order of $(Q/2m_{\pi})^2\approx 9$\%. A somewhat larger 
value is obtained by using a mean $\Lambda N$ pair breakup energy in $_{
\Lambda}^5$He, $B_{\Lambda N}$=($B_{\Lambda}$+$B_N$)/2=12.1~MeV, to estimate 
$p_{\Lambda N}\approx\sqrt{2\mu_{\Lambda N}B_{\Lambda N}}$ in light $\Lambda$ 
hypernuclei. This yields $p_{\Lambda N}\approx 111$~MeV/c and $(p_{\Lambda N}
/2m_{\pi})^2\approx 0.16$. With past \nopieft~$\Lambda$ hypernuclear 
applications limited to $A=3$ systems~\cite{hammer02,ARO15}, ours is the first 
comprehensive application of \nopieft~to the full hypernuclear $s$ shell. 

As shown in this Letter, our few-body SVM calculations of light $\Lambda$ 
hypernuclei in the \nopieft~approach largely resolve the overbinding problem 
for $_{\Lambda}^5$He to the accuracy expected at LO. Below, we expand briefly 
on the \nopieft~approach, its input, and the SVM few-body calculations applied 
in the present work to light nuclei and hypernuclei. Possible consequences of 
resolving the hypernuclear overbinding problem in light hypernuclei and 
extensions to heavier systems are discussed in the concluding paragraphs. 
\newline 
\noindent
{\bf Application of \nopieft~to $\Lambda$ hypernuclei.}~~ 
Hadronic systems consisting of neutrons, protons, and $\Lambda$-hyperons 
are described in $\nopieft$ by a Lagrangian density 
\begin{equation} 
\Lag = N^\dagger (i\partial_0+\frac{\nabla^2}{2 M_N})N 
 + \Lambda^\dagger (i\partial_0+\frac{\nabla^2}{2 M_\Lambda})\Lambda 
 + \Lag_{2B}+\Lag_{3B}+\ldots\,,
\label{eq:Lag} 
\end{equation} 
where $N$ and $\Lambda$ are nucleon and $\Lambda$-hyperon fields, 
respectively, and $\Lag_{2B},\Lag_{3B},\ldots $ are two-body, three-body, and, 
in general, $n$-body interaction terms. The interaction terms are composed of 
$N,\Lambda$ fields and their derivatives subject to symmetry constraints that 
$\Lag$ is scalar and isoscalar and to a power counting that orders them 
according to their importance. At LO, the Lagrangian contains only contact 
two-body and three-body $s$-wave interaction terms; i.e., $\Lag_{2B}$ and 
$\Lag_{3B}$ are the sum of all possible $N,\Lambda$ field combinations, with 
no derivatives, that create an $s$-wave projection operator. Thus, there is 
a one-to-one correspondence between LO interaction terms and all possible $NN,
N\Lambda,\Lambda\Lambda$ and $NNN,NN\Lambda,\ldots$ $s$-wave states. Each 
of these terms is associated with its own low-energy constant (LEC). In the 
present work we focus on single-$\Lambda$ hypernuclei and, hence, ignore all 
terms in $\Lag$ containing more than one $\Lambda^\dagger_\alpha\Lambda_\beta$ 
field pair. 

Momentum-dependent interaction terms, such as tensor or spin-orbit, appear at 
subleading order in $\nopieft$ power counting~\cite{Kol99}. In particular, the 
long-range $\Lambda N$ tensor force induced by a $\Lambda N \to \Sigma N$ OPE 
transition followed by a $\Sigma N \to \Lambda N$ OPE transition is expected 
to be weak because this two-pion exchange mechanism is dominated by its 
central $S\to D\to S$ component, which is partially absorbed in the $\Lambda 
N$ and $\Lambda NN$ LO contact LECs. Short-range $K$ and $K^\ast$ exchanges 
add a rather weak direct $\Lambda N$ tensor force~\cite{GSD71,MGDD85}, 
as also deduced from several observed $p$-shell $\Lambda$ hypernuclear 
spectra~\cite{Mil12}. 

The contact interactions of the Lagrangian $\Lag$ are regularized by 
introducing a local Gaussian regulator with momentum cut-off $\lambda$ 
(see, e.g.,~\cite{Bazak16}): 
\begin{equation} 
\delta_\lambda(\rvec)=\left(\frac{\lambda}{2\sqrt{\pi}}\right)^3\,
\exp \left(-{\frac{\lambda^2}{4}}\rvec^2\right) 
\label{eq:gaussian} 
\end{equation} 
that smears contact terms over distances~$\sim\lambda^{-1}$, becoming a Dirac 
$\delta^{(3)}(\rvec)$ in the limit $\lambda\to\infty$. The cut-off parameter 
$\lambda$ may be viewed as a scale parameter with respect to typical values 
of momenta $Q$. To make observables independent of specific values of 
$\lambda$, the LECs must be properly renormalized. Truncating \nopieft~at LO 
and using values of $\lambda$ higher than the breakup scale of the theory 
(here $\approx$2$m_{\pi}$), observables acquire a residual dependence 
$O(Q/\lambda)$ which diminishes with increasing $\lambda$. 

\begin{table}[ht!] 
\caption{Input scattering lengths (in fm) used to fit \nopieft~two-body LECs; 
see the text.} 
\begin{ruledtabular} 
\begin{tabular}{cccccc} 
$YN$ model & Ref. & $a_s(NN)$ & $a_s(\Lambda N)$ & $a_t(\Lambda N)$ & 
${\bar a}_{\Lambda N}$  \\
\hline 
Alexander[A] & \cite{Alex68} & $-$23.72 & $-$1.8 & $-$1.6 & $-$1.65  \\
Alexander[B] & \cite{Alex68} & $-$18.63 & $-$1.8 & $-$1.6 & $-$1.65  \\
NSC97f       & \cite{NSC97}  & $-$18.63 & $-$2.60 & $-$1.71 & $-$1.93  \\ 
$\chi$EFT(LO) & \cite{Polinder06} & $-$18.63 & $-$1.91 & $-$1.23 & $-$1.40  \\ 
$\chi$EFT(NLO) & \cite{Haiden13} & $-$18.63 & $-$2.91 & $-$1.54 & $-$1.88  \\ 
\end{tabular} 
\end{ruledtabular} 
\label{tab:LN} 
\end{table} 

The resulting LO two-body interaction is given by 
\begin{equation}  
V_{2B} = \sum_{IS}\,C_{\lambda}^{IS} \sum_{i<j} {\cal P}_{IS}(ij)
          \delta_\lambda(\rvec_{ij}), 
\label{eq:V2} 
\end{equation} 
where ${\cal P}_{IS}$ are projection operators on $NN,\Lambda N$ pairs with 
isospin $I$ and spin $S$ and $C_{\lambda}^{IS}$ are LECs, fixed by fitting 
to low-energy two-body observables, e.g., to the corresponding $NN$ and 
$\Lambda N$ scattering lengths. In the present work the $NN$ $IS$=01 LEC is 
fitted to the deuteron binding energy, hardly affecting the results obtained 
alternatively by fitting to the $IS$=01 scattering length. The scattering 
lengths used to fit the LECs are listed in Table~\ref{tab:LN}. For $IS$=10, 
two choices of a charge-independent $NN$ spin-singlet scattering length, 
[A] and [B], were made for comparison~\cite{Miller06}. For $\Lambda N$ 
scattering lengths we used best-fit values derived from the low-energy 
$\Lambda p$ spin-averaged scattering cross sections measured by Alexander 
et al.~\cite{Alex68}, assuming charge symmetry, and also values from several 
listed $YN$ interaction models. These choices suggest a $^1S_0$ $\Lambda N$ 
interaction stronger than in $^3S_1$, spanning a broad range of possible 
$\Lambda N$ spin dependence. Also listed are values of the spin-averaged 
$\Lambda N$ scattering length $\bar{a}$=(3$a_t$+$a_s$)/4, with approximately 
$\pm$16\% spread about the best-fit value $-$1.65~fm from Ref.~\cite{Alex68}, 
reflecting the model dependence of fitting \textit{all} low-energy $YN$ 
scattering and reaction cross section data~\cite{extrap}. 

\begin{figure*}[ht!] 
\centering 
\includegraphics[width=0.48\linewidth]{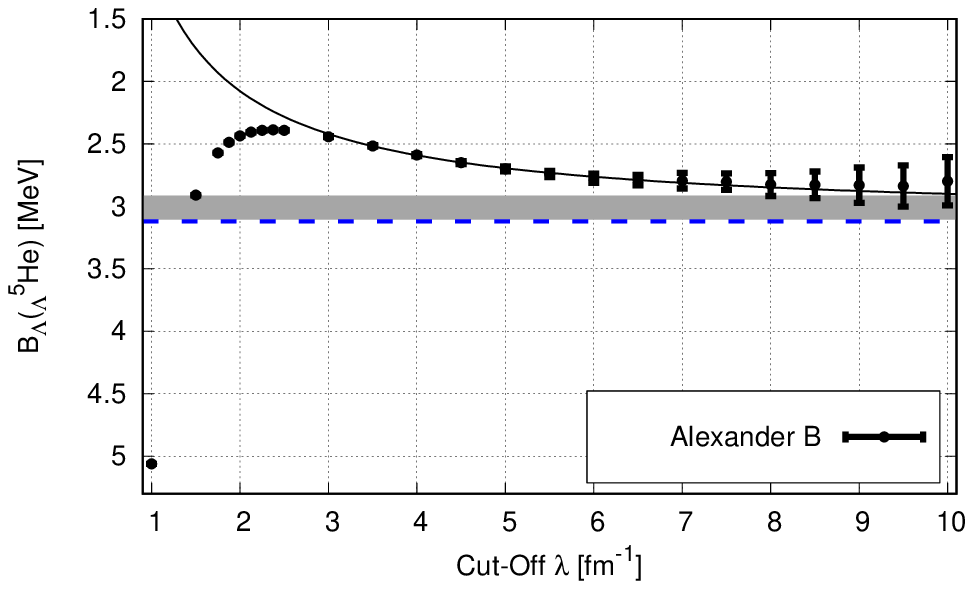} 
\includegraphics[width=0.48\linewidth]{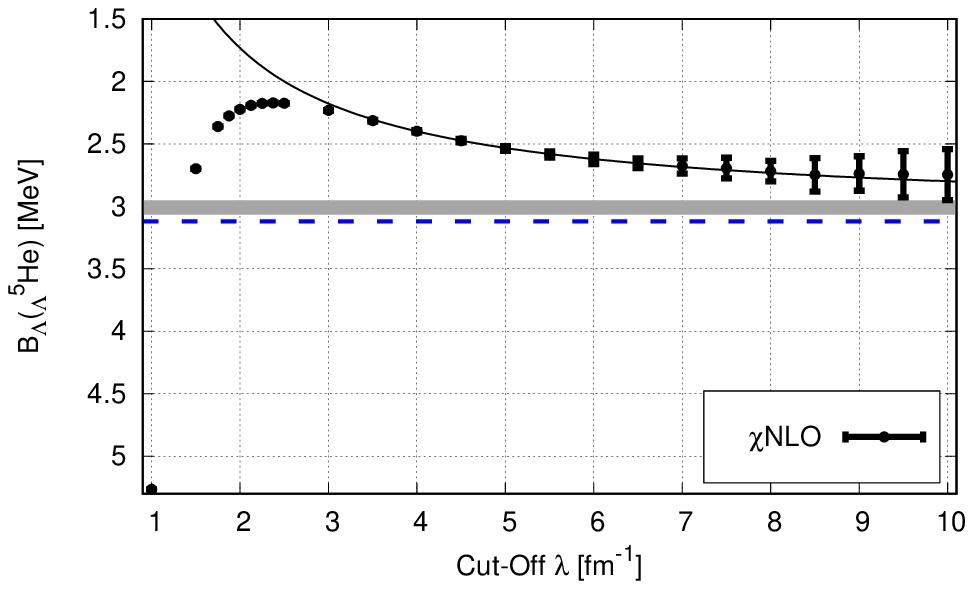} 
\caption{$B_{\Lambda}(_\Lambda^5$He) (MeV) as a function of the cut-off 
$\lambda$ (fm$^{-1}$) in LO \nopieft~calculations with $\Lambda N$ 
scattering-length input listed in Table~\ref{tab:LN}. Solid lines mark 
a two-parameter fit $a+b/\lambda$, starting from $\lambda=4$~fm$^{-1}$. Gray 
horizontal bands mark $\lambda\to\infty$ extrapolation uncertainties. Dashed 
horizontal lines mark the value $B_{\Lambda}^{\rm exp}(_{\Lambda}^5
$He)=3.12$\pm$0.02~MeV.} 
\label{fig:L5He} 
\end{figure*} 

The LO three-body interaction consists of a single $NNN$ term associated 
with the $IS=\frac{1}{2}\frac{1}{2}$ channel and of three $\Lambda NN$ terms 
associated with the $IS=0\frac{1}{2},1\frac{1}{2},0\frac{3}{2}$ $s$-wave 
configurations. The explicit form of the three-body $NNN$ potential is given 
by
\begin{equation} 
V_{NNN} = D_{\lambda}^{\half\half}\sum_{i<j<k}{\cal Q}_{\half\half}(ijk)
\left(\sum_{\rm cyc}\delta_\lambda(\rvec_{ik})\, \delta_\lambda(\rvec_{jk})
\right),
\label{eq:V3NNN}
\end{equation} 
where the first sum runs over all $NNN$ triplets. The three-body $\Lambda NN$ 
potential is given by 
\begin{equation} 
V_{\Lambda NN} = \sum_{IS}\,D_{\lambda}^{IS} \sum_{i<j}\,
{\cal Q}_{IS}(ij\Lambda)\,\delta_\lambda(\rvec_{i\Lambda})\,
\delta_\lambda(\rvec_{j\Lambda}),
\label{eq:V3NNL} 
\end{equation} 
where the second sum runs over all $NN$ pairs. 
In Eqs.~(\ref{eq:V3NNN}) and (\ref{eq:V3NNL}), ${\cal Q}_{IS}$ are 
projection operators on baryon triplets with isospin $I$ and spin $S$, 
and $D_{\lambda}^{IS}$ are LECs. 

There are four three-body LECs, a pure $NNN$ LEC $D_{\lambda}^{\half\half}$ 
fitted to $B(^3$H) and three $\Lambda NN$ LECs associated with the three 
possible $s$-wave $\Lambda NN$ systems. Because only $_{\Lambda}^3$H($I$=0,$
J^P$=${\frac{1}{2}}^+$) is known to be bound, we have fitted these LECs 
instead to the three $B_{\Lambda}$ values available (disregarding CSB) 
for $A\leq 4$: $_\Lambda^3$H($I$=0,$J^P$=${\frac{1}{2}}^+$) for 
$D_{\lambda}^{0\frac{1}{2}}$, $_\Lambda^4$H$_{\rm gs}$($I$=$\frac{1}{2}
$,$J^P$=$0^+$) subsequently for $D_{\lambda}^{1\frac{1}{2}}$, and finally 
$_\Lambda^4$H$_{\rm exc}$($I$=$\frac{1}{2}$,$J^P$=$1^+$) for 
$D_{\lambda}^{0\frac{3}{2}}$. Altogether, eight LECs at LO are constrained 
by few-body nuclear and hypernuclear data, to be subsequently used in 
calculations of $^4$He and $_{\Lambda}^5$He. 
\newline 
\noindent 
{\bf Stochastic variational method (SVM).}~~
To solve the $A$-body Schr\"{o}dinger equation, the wave function $\Psi$ 
is expanded on a correlated Gaussian basis. Introducing a vector 
$\xvec=(\xvec_1,\xvec_2,\ldots,\xvec_{A-1})$ of Jacobi vectors 
$\xvec_j$, $j$=$1,2,\ldots,A-1$, we may write $\Psi$ as 
\begin{equation} 
\Psi = \sum_k c_k \hat{\cal A} \left\{\chi_{S M}^k\xi_{I I_z}^k
        {\exp\left(-{\frac{1}{2}}{\xvec}^T A_k{\xvec}\right)} \right\} \;, 
\label{eq:Psi} 
\end{equation} 
where the operator $\hat{\cal A}$ antisymmetrizes over nucleons. 
In Eq.~(\ref{eq:Psi}) the basis states are defined by the real, symmetric and 
positive-definite $(A-1)\times(A-1)$ matrix $A_k$, together with the spin and 
isospin functions $\chi_S$ and $\xi_I$. Once these are chosen, the linear 
variational parameters $c_k$ are obtained through diagonalization of the 
Hamiltonian matrix. The matrix $A_k$ introduces $A(A-1)/2$ nonlinear 
variational parameters which are chosen stochastically, hence the name 
SVM. For a comprehensive review, see Ref.~\cite{SVa98}. For the specific 
calculation of the three-body interaction matrix elements, 
see Ref.~\cite{Bazak16}.  
\newline 
\noindent 
{\bf Results and discussion.}~~
The \nopieft~approach with two-body and three-body regulated contact terms 
defined by Eqs.~(\ref{eq:V2})--(\ref{eq:V3NNL}) was applied in SVM few-body 
calculations as outlined above to the $s$-shell nuclei and hypernuclei using 
the $\Lambda N$ scattering-length combinations listed in Table~\ref{tab:LN}. 
The calculated ${_\Lambda^5}$He binding energy $B({_\Lambda^5}$He) along with 
$B(^4$He) are found to depend only moderately on $\lambda$, for $\lambda
\gtrsim 2$~fm$^{-1}$, exhibiting renormalization scale invariance in the 
limit $\lambda\to\infty$. Using $a_s(NN)$=$-$18.63~fm, we obtain in this 
limit $B(^4$He)$\to$29.2$\pm$0.5~MeV, which compares well with $B_{\rm exp}
(^4$He)=28.3~MeV, given that our $\nopieft$ is truncated at LO and considering 
that the suppressed Coulomb force is expected to reduce $B(^4$He) further by 
roughly 1~MeV. The binding energies $B(^4$He) calculated for the other choice, 
$a_s(NN)$=$-$23.72~fm, differ by less than 0.4~MeV and agree with those 
calculated recently in Ref.~\cite{CLP17}. 

With $B(^4$He) and $B(_\Lambda^5$He) computed, we show in Fig.~\ref{fig:L5He} 
the resulting $\Lambda$ separation energy values $B_{\Lambda}(_\Lambda^5$He) 
as a function of the cut-off $\lambda$ for the $\Lambda N$ scattering-length 
versions Alexander[B] and $\chi$NLO listed in Table~\ref{tab:LN}. The results 
shown for Alexander[B] agree to a level of 1\% with those (not shown) for 
Alexander[A]; both versions differ only in their $^1S_0$ $NN$ input. The 
dependence of the calculated $B_{\Lambda}(_\Lambda^5$He) values on $\lambda$ 
is similar in all versions, switching from about 2--3~MeV overbinding at 
$\lambda$=1~fm$^{-1}$ to less than 1~MeV underbinding between $\lambda
$=2 and 3~fm$^{-1}$, and smoothly varying beyond, approaching a finite limit 
at $\lambda\to\infty$. Renormalization scale invariance implies that $B_{
\Lambda}(_\Lambda^5$He) should be considered in this limit. However, it may 
be argued that, when the cut-off value $\lambda$ matches the EFT breakup 
scale, higher-order terms such as effective-range corrections are absorbed 
into the LECs. A reasonable choice of {\it finite} cut-off values in the 
present case is between $\lambda\approx 1.5$~fm$^{-1}$, which marks the 
\nopieft~breakup scale of 2$m_{\pi}$, and 4~fm$^{-1}$, beginning at which 
the detailed dynamics of vector-meson exchanges may require attention. 
In the following we compare the finite versus infinite options for $\lambda$.  

\begin{table}[htb] 
\caption{$B_{\Lambda}(_\Lambda^5$He) values (MeV) in LO \nopieft~calculations 
for several $\Lambda N$ scattering-length versions from Table~\ref{tab:LN}. 
The uncertainties listed for cut-off $\lambda$=4~fm$^{-1}$ are due to 
subtracting $B(^4$He) from $B(_\Lambda^5$He), whereas those for $\lambda\to
\infty$ are mostly from extrapolation, with fitting uncertainties $\lesssim 
10$~keV.} 
\begin{ruledtabular}
\begin{tabular}{ccccc} 
$\lambda$ (fm$^{-1}$) & Alexander[B] & NSC97f & $\chi$LO 
& $\chi$NLO \\ 
\hline 
4 & 2.59(3) & 2.32(3) & 2.99(3) & 2.40(3) \\ 
$\to\infty$  & 3.01(10) & 2.74(11) & 3.96(08) & 3.01(06) \\ 
\end{tabular} 
\end{ruledtabular} 
\label{tab:B5}  
\end{table} 

Calculated values of $B_{\Lambda}(_\Lambda^5$He) are listed in 
Table~\ref{tab:B5} for $\lambda$=4~fm$^{-1}$ and as extrapolated to 
$\lambda\to \infty$. To extrapolate to $\lambda\to\infty$, the calculated 
$B(\lambda)$ values can be fitted by a power series in the small parameter 
$Q/\lambda$:
\begin{equation} 
\frac{B(\lambda)}{B(\infty)}=\left[1+\alpha\frac{Q}{\lambda}
            +\beta\left(\frac{Q}{\lambda}\right)^2
            +\gamma\left(\frac{Q}{\lambda}\right)^3
            +\ldots\right].  
\label{eq:extrap} 
\end{equation} 
The extrapolation uncertainties listed in Table~\ref{tab:B5} for the 
asymptotic values $B_{\Lambda}(\lambda\to\infty)$ were derived by comparing 
two- and three-parameter fits of this form. These uncertainties are also 
shown as gray bands in Fig.~\ref{fig:L5He} above. The table demonstrates 
how $\Lambda N$ version $\chi$LO, of all versions, is close to reproducing 
$B_{\Lambda}^{\rm exp}(_\Lambda^5$He) for $\lambda$=4~fm$^{-1}$, whereas 
versions Alexander[B] and $\chi$NLO (see also Fig.~\ref{fig:L5He}) do so 
only in the limit $\lambda\to\infty$. 

The sign and size of the three-body contributions play a crucial role 
in understanding the cut-off $\lambda$ dependence of the calculated 
$B_{\Lambda}(_\Lambda^5$He). The nuclear $NNN$ term first changes from weak 
attraction at $\lambda$=1~fm$^{-1}$ in $^3$H and $^4$He, similar to that 
required in phenomenological models~\cite{NKG00}, to strong repulsion at 
$\lambda$=2~fm$^{-1}$, which reaches maximal values around $\lambda$=4~fm$^{-1}
$. However, for larger values of $\lambda$ it decreases slowly. The $\Lambda 
NN$ contribution follows a similar trend, but it is weaker than the $NNN$ 
contribution by a factor of roughly 3 when repulsive. The transition of the 
three-body contributions from long-range weak attraction to relatively strong 
repulsion for short-range interactions is correlated with the transition seen 
in Fig.~\ref{fig:L5He} from strongly overbinding $^5_\Lambda$He to weakly 
underbinding it. We note that for $\lambda\gtrsim 1.5$~fm$^{-1}$ all of the 
three $\Lambda NN$ components are repulsive, as required to avoid Thomas 
collapse, imposing thereby some constraints on the $\Lambda NN$ LECs. 

Finally, using the $\nopieft$ LECs derived here to evaluate $B_{\Lambda}$ in 
symmetric nuclear matter (SNM), we have found within a simple Fermi gas model 
that for version Alexander[B], for example, $B_{\Lambda}({\rm SNM})\leq 27
$~MeV at nuclear saturation density, $\rho_A = 0.16$~fm$^{-3}$, for any 
cut-off value $\lambda$. Although this value is only a lower bound 
on the binding energy of $\Lambda$ in SNM, the acceptable value being 
$\approx$30~MeV~\cite{GHM16}, it is encouraging that our \nopieft~does 
not lead to excessive binding. This calls for more rigorous evaluations of 
$B_{\Lambda}({\rm SNM})$ using perhaps advanced Monte Carlo variational 
techniques. 
\newline
\noindent 
{\bf Summary and outlook.}~~ 
The present work was motivated by the 1--3 MeV persistent overbinding of 
${_{\Lambda}^5}$He in most of the few-body calculations reported to date, 
including recent LO EFT model calculations~\cite{WR18}. To this end, we have 
applied the $\nopieft$ approach at LO to $s$-shell $\Lambda$ hypernuclei 
within precise few-body SVM calculations, extending recent $\nopieft$ studies 
of light nuclei~\cite{BCG15,KBG15,CLP17,KPDB17}. This required five LECs 
at LO: two $\Lambda N$ LECs, related here to spin-triplet and spin-singlet 
$\Lambda N$ scattering lengths in several $\Lambda N$ interaction models, and 
three $\Lambda NN$ LECs fitted to the three available $B_{\Lambda}$ values 
in the $A$=3,4 hypernuclei. With these five fitted LECs, for each of the 
momentum scale parameters $\lambda$ chosen, the $\Lambda$ separation energy 
$B_{\Lambda}({_{\Lambda}^5}$He) was evaluated. Our main finding is that, 
while ${_{\Lambda}^5}$He is overbound indeed by up to 3~MeV for relatively 
long-range $\Lambda N$ and $\Lambda NN$ interactions, say at $\lambda\sim 
1$~fm$^{-1}$, it quickly becomes underbound by less than 1~MeV for 
$\lambda\sim 2-3$~fm$^{-1}$. For most of the $\Lambda N$ scattering-length 
versions studied here, $B_{\Lambda}^{\rm calc}({_{\Lambda}^5}$He) approaches 
slowly in the limit $\lambda\to\infty$ the value $B_{\Lambda}^{\rm exp}
(_{\Lambda}^5$He)=3.12$\pm$0.02~MeV, notably for version Alexander[B] 
derived in a model independent way directly from experiment. 

Having largely resolved the overbinding problem in light $\Lambda$ 
hypernuclei, it would be interesting in future work to study possible 
implications of the strong three-body $\Lambda NN$ interactions found here 
to other problems that involve hyperons in nuclear and neutron-star matter. 
To be more specific, we make the following observations: 
\newline  
(i) Other than the $s$-shell hypernuclei studied in the present work, 
$p$-shell hypernuclei offer a well-studied range of mass numbers $6\leq A\leq 
16$ both experimentally and theoretically~\cite{GHM16}. Recent $\chi$EFT LO 
calculations~\cite{WR16} using induced $YNN$ repulsive contributions suggest 
that the $s$-shell overbinding problem extends to the $p$ shell. In contrast, 
shell-model studies~\cite{Mil12} reproduce satisfactorily $p$-shell 
ground-state $B_{\Lambda}$ values, essentially by using $B_{\Lambda}^{\rm exp}
(_{\Lambda}^5$He) for input, except for the relatively large difference of 
about 1.8~MeV between $B_{\Lambda}({_{\Lambda}^9}$Li) and $B_{\Lambda}({_{
\Lambda}^9}$Be). In fact, it was noted long ago that strongly repulsive 
$\Lambda NN$ terms could settle it~\cite{gal67}. It would be interesting 
to apply our derived $\Lambda NN$ interaction terms in future shell-model 
calculations. 
\newline 
(ii) The \nopieft~Hamiltonian derived here includes already at LO repulsive 
$\Lambda NN$ terms which are qualitatively as strong as those used by 
Lonardoni, Pederiva and Gandolfi~\cite{LPG14} to resolve the hyperon puzzle 
\cite{LLGP15}. It would be interesting then to apply our $\Lambda N$+$\Lambda 
NN$ interaction terms in state-of-the-art neutron-star matter calculations 
to see whether or not their suggested resolution of the hyperon puzzle is 
sufficiently robust. 

We hope to discuss in greater detail some of these issues in forthcoming 
studies. 
\newline 
\newline 
The work of L.C. and N.B. was supported by the Pazy Foundation and by the 
Israel Science Foundation Grant No. 1308/16.

\end{document}